\documentclass[superscriptaddress,twocolumn,aps,amsmath,amssymb]{revtex4}
\usepackage{amsmath, amssymb, amsthm, alltt, setspace,bbm}
\usepackage[bookmarks=true, pdfpagemode = None, pdfstartview = FitH,colorlinks =
true, citecolor = black, linkcolor = black, urlcolor=black]{hyperref}
\usepackage{Karnaugh}
\usepackage{graphicx}
\usepackage{amssymb}

\newcommand{\bra}[1]{\langle#1|}
\newcommand{\ket}[1]{|#1\rangle}

\newcommand{\iseq}{\stackrel{\mathrm{?}}{=}}

\newcommand{\spn}{\operatorname{span}}

\usepackage{amsthm}
\theoremstyle{plain}
\newtheorem{theorem}{Theorem}[section]
\newtheorem{lemma}[theorem]{Lemma}

\theoremstyle{definition}

\newcommand{\qedsymb}{\hfill{\rule{2mm}{2mm}}}

\theoremstyle{remark}

\newcommand{\QMA}{{\sf QMA}}

\newcommand{\NP}{{\sf{NP}}}

\def\LH{{\sc local Hamiltonian}}
\def\2LH{{\sc $2$-local Hamiltonian}}
\def\3LH{{\sc $3$-local Hamiltonian}}
\def\5LH{{\sc $5$-local Hamiltonian}}

\def\3SAT{{\sc $3$-SAT}}

\def\AND{{\sf AND}}
\def\OR{{\sf OR}}

\def\XOR{{\sf XOR}}
\def\EQV{{\sf EQV}}

\newcommand{\bydef}{\stackrel{\mathrm{def}}{=}}
\begin{document}
\title{Non-perturbative k-body to two-body commuting conversion Hamiltonians and embedding problem instances into Ising spins}%
\author{J.D. Biamonte}\email{jacob.biamonte@comlab.ox.ac.uk}\affiliation{Oxford University Computing Laboratory, \\Wolfson Building, Parks Road, Oxford, OX1 3QD, United Kingdom.}\affiliation{Department of Chemistry and Chemical Biology, Harvard University, \\12 Oxford St., Cambridge, MA, 02138, USA.}

\begin{abstract}

An algebraic method has been developed which allows one to engineer
several energy levels including the low-energy subspace of interacting spin systems.  By introducing
ancillary qubits, this approach allows k-body interactions to be captured exactly using 2-body Hamiltonians.  Our method works when all terms in the Hamiltonian share the same basis and has no dependence on
perturbation theory or the associated large spectral gap.  Our methods allow problem instance solutions to be embedded into the ground energy state of Ising spin systems.  Adiabatic evolution might then be used to place a computational system into it's ground state.


\end{abstract}
\maketitle


This work considers an important problem.  Given a Hamiltonian comprised solely of 1-body and 2-body terms, from this Hamiltonian, and with the aid of ancillary qubits, is it possible to construct the ground states of a Hamiltonian containing k-body terms with respect to a suitable subspace?  In both the classical and quantum cases, this problem is particularly important when considering the physical complexity of interacting spin systems evolving into their lowest energy configuration~\cite{KGV83,BBRA99,FGGS00,RO05} or the equivalent computational task of determining the ground state~\cite{Bar82,KSV02}.

The ground state energy problem has long been considered in the realm of classical complexity theory with well known results appearing in work such as~\cite{Bar82,KGV83}.  The extension to quantum complexity classes was prompted when Kitaev~\cite{KSV02}, inspired by ideas from Feynman~\cite{Fey82}, showed that the ground state energy problem of the 5-local (that is, 5-body) random field quantum spin model was complete for the quantum analogue of the class \NP{}.  Thus it was shown that 5-\LH{} was \QMA{}-complete and the quest to determine the complexity of various spin models began~\cite{KR03,NM06,KKR06,OT06,BL07,LCV06,AGK07,Ira07}.

Ideas from the theory of quantum computation have also led to the use of ground state properties of quantum systems for computation~\cite{FGGS00,FGG02,AvDK+05}.  This is known as the adiabatic model of quantum computation~\cite{FGGS00,FGG02} --- in which a driving Hamiltonian is slowly replaced, most often with a commuting Hamiltonian with a ground state spin configuration representing a problem instance solution.


At the heart of the construction of the \QMA{}-completeness proofs lies the development of methods to engineer low-energy effective Hamiltonians, which approximate k-body interactions, using at most 2-body terms~\cite{KKR06,OT06,BL07}.  To date, all known methods require the introduction of a large spectral gap, where the magnitude of the gap improves only an approximate low-energy effective Hamiltonian.  It would be desirable if one could \emph{i.)} remove the spectral gap dependence by capturing the low-energy effective subspace exactly and  \emph{ii.)} develop a systematic method to engineer multiple energy subspaces, including any ground state.

The present paper addresses both of these problems.  Somewhat surprisingly, it is possible to remove dependence on the large spectral gap by allowing the state of the ancillary mediator qubits (facilitating the coupling) to \emph{follow} the state of the qubits being coupled.  In application, care is taken to ensure that the active role of the mediator qubits is appropriate for any given application.  In many cases, this new approach allows ground states of k-body interactions to be captured exactly using 2-body interactions; under the restriction that all terms in the Hamiltonian share the same basis.

\textbf{Structure:} The remainder of this paper begins with a short introduction, followed by  $\S$~\ref{sec:3local}, which explains how the ground states of 3-body Hamiltonians can be used to embed any Boolean function (and for that matter, any switching circuit).  $\S$~\ref{sec:2local} reduces the 3-local Hamiltonians used in $\S$~\ref{sec:3local} to the case of 2-local Hamiltonians:  In addition, we prove Theorem~\ref{theorem:klocalboolean}, which states the existence of an efficient method to construct Hamiltonians that simulate Boolean functions containing k-variable couplings (i.e. $x_1\wedge x_2\wedge\cdots \wedge x_k$). In $\S$~\ref{sec:novelgadget} we construct 2-body Hamiltonians that exactly capture the ground space of k-body Hamiltonians of the form $J\sigma_1 \otimes\sigma_2\otimes\cdots \otimes\sigma_k$.  $\S$~\ref{sec:novelgadget} also contains a proof of Theorem~\ref{theorem:manylevels}, which states the existence of a method to construct several energy subspaces of a given Hamiltonian --- a necessity for certain applications.

In addition to the main body of the present paper, Appendix~\ref{appendix:projection} presents a proof of a tailored variant of the projection Lemma~\cite{KR03,AvDK+05,KKR06}.  This is followed by Appendix~\ref{appendix:kmap} which explains Karnaugh maps --- key to an algebraic reduction method relied on during several derivations.  We make use of standard quantum computing notation and background information~\cite{KSV02,KR03} as well as that for discrete functions and circuits~\cite{Weg87,BH02}.


\section{Introduction}\label{sec:label}


Let us represent an Ising spin with index $i$ by the variable $s_i\in\{+1,-1\}$.  One could also represent variable $s_i$ in terms of binary variable $x_i\in\{0,1\}$ as $s_i=1-2x_i$, which we will denote as $\ket{x_i}$.  A single spin system can be acted on by linear combinations of operators taken from the set $\{\openone , \pm\sigma\}$, where the identity operator ($\openone $) can be scaled to ensure positive-semidefiniteness and the operator $\sigma$ has eigenvectors $\ket{0}$ and $\ket{1}$ with respective eigenvalues $+1$ and $-1$.  The energy levels of the Hamiltonian operator $\frac{1}{2}(\openone+\sigma_i)$ $\left(\frac{1}{2}(\openone-\sigma_i)~\text{respectively} \right)$ corresponding to the states $\ket{0}$ and $\ket{1}$ are $1$ and $0$ ($0$ and $1$ respectively). Addition of the operator $\frac{1}{2}(\openone+\sigma_i)$ $\left(\frac{1}{2}(\openone-\sigma_i)\right)$ adds an energy penalty to the state $\ket{0}$ ($\ket{1}$) and can be thought of as negation (assignment) of variable $x_i$.

In the case of two Ising spins, a complete basis of configurations are $\ket{00}, \ket{01}, \ket{10}$ and $\ket{11}$.  Let us add scaled sums of a coupling term to our Hamiltonian: $\pm\sigma_i\sigma_j$.  One can think of adding the operator $\frac{1}{2}(\openone-\sigma_i\sigma_j)$ as a logical equality operation (i.e. the characteristic function $x_i\Leftrightarrow x_j$ is true) and the operator $\frac{1}{2}(\openone+\sigma_i\sigma_j)$ as a logical inequality operation (i.e. $x_i\nLeftrightarrow x_j$ is true) between spins.  For example, assume we act on a dual spin system with the Hamiltonian for inequality:  the ground space is in $\spn\{\ket{01},\ket{10}\}$, so any vector that corresponds to two spin variables being equal (e.g. $\spn\{\ket{11},\ket{00}\}\bydef\spn\{\ket{x}\ket{y}|x=y, \forall x,y\in\{0,1\}\}$) receives an energy penalty.

We have shown how to set single spin variables, and how to apply equality and inequality operations between two spins.  These operations, however, do not form a convenient logical system~\footnote{It is known that finding the ground state of Hamiltonians formed from simple sums of the inequality operator $x_i\nLeftrightarrow x_j$ is \NP{}-complete on a planar graph~\cite{Bar82}.}.  This will be done next, in $\S$~\ref{sec:3local} and~\ref{sec:2local}, by defining Hamiltonians with ground state spin configurations representing logical operations such as the \AND{} ($\wedge$) gate, the \OR{} ($\vee$) gate, etc.  We know that these dual arity operations require at least three spins as $x_i\square x_j=z_\star$.  What we need is to find a way to set the low-energy subspace of three spins $s_i$, $s_j$ and $z_\star$ to be, for instance, the logical \AND{} of the spins $s_i\wedge s_j=z_\star$.  This assignment turns out to be possible working in the energy basis of a Hamiltonian equipped with a commuting local field and coupling term, such as an Ising Hamiltonian~\footnote{It is understood that a term in a Hamiltonian such as $\sigma_i\sigma_j$ is the operator $\sigma$ acting on the $i^{th}$ and $j^{th}$ qubit with the omitted identity operator acting on the rest of the Hilbert space.  The tensor product symbol ($\otimes$) is omitted between operators.}:
\begin{equation}\label{eqn:H1}
  H_{\text{Ising}} = \sum_ih_i \sigma_i+\sum_{\langle i,j\rangle}J_{ij} \sigma_i\sigma_j.
\end{equation}
Impressive demonstrations using qubits based on Josephson junctions~\cite{Jgate,flux_qubit,2006cond.mat..8253H} make an adiabatic~\cite{FGGS00,FGG02} realization of ground state logic gates using variants of the Hamiltonian (\ref{eqn:H1}) a foreseeable possibility. 


\section{Ground state spin logic}\label{sec:3local}
Consider some Hamiltonian $H$ acting on a Hilbert space $\mathcal{H}$ that is a sum of the vectors spanned by the subspace ${\cal L}$ and the orthogonal component of ${\cal L}$ written as ${\cal L}^\perp$, thus $\mathcal{H} =  {\cal L}^\perp + {\cal L}$.  The lowest eigenvalue of $  H$ will be denoted as $\lambda(H)$. Now let $\Pi_{\cal L}\bydef (\openone-{\cal L})$ be defined as a projector onto ${\cal L}$.  Then $\Pi_{\cal L}H \Pi_{\cal L}$ is the restriction of $H$ to the subspace ${\cal L}$ --- let us write this restriction as $H|_{\cal L}$.

To develop the logic, consider the Hamiltonian $H_\text{prop}$ such that $H_\text{prop}|_{\cal L}=0$ and $H_\text{prop}|_{\cal L^\perp}\geq\delta~(>2\|  H_\text{in}\|)$ where $H_\text{in}$ is a perturbation later used to set the circuits inputs, the norm $\|\cdot\|$ is the magnitude of the Hamiltonians largest eivenvalue and $\delta$ is the spectral gap between the ${\cal L^\perp}$ and ${\cal L}$ subspaces.  We are faced with the task of ensuring that $  H_\text{prop}|_{\cal L}$ is a zero eigenspace when ${\cal L}$ spans the truth table of the logical operation of interest (e.g. ${\cal L}=\spn\{\ket{x_1}\ket{x_2}\ket{x_1\square x_2}|\forall x_1,x_2\in\{0,1\}\}$).  Let ${\cal L}$ be the low-energy subspace representing the truth table in the binary observables.  Explicitly, in the case of logical \AND{}, ${\cal L}=\spn\{\ket{000},\ket{010},\ket{100},\ket{111}\}$ (ordered $\ket{x_1 x_2}\ket{z_\star}$, where $z_\star = x_1\wedge x_2$), which is a zero eigenspace of $H_\text{prop}$ and ${\cal L^\perp}=\spn\{\ket{001},\ket{011},\ket{101},\ket{110}\}$ ($z_\star \neq x_1\wedge x_2$) will be all eigenspaces of at least $\delta$~\footnote{A simplistic Hamiltonian with vectors in the ground space ${\cal L}$ corresponding to logical \AND{}, that is ${\cal L}=\spn\{\ket{000},\ket{010},\ket{100},\ket{111}\}$ (ordered $\ket{x_1 x_2}\ket{z_\star}$, where $z_\star = x_1\wedge x_2$), has the form: $H=\delta(\openone  -\ket{000}\bra{000}-\ket{010}\bra{010}-\ket{100}\bra{100}-\ket{111}\bra{111})$}.

\begin{figure}[t]\label{fig:cc}
\center
\includegraphics[width=5cm]{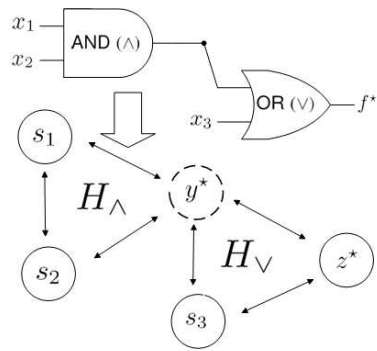}
\caption{Illustrating the mapping between circuits (with boolean variables $x_i$) and spins ($s_i$) for the example given in (\ref{eqn:e1}).  One can use any number of methods to embed logical networks~\cite{Weg87} into the ground space of Hamiltonians.}
\end{figure}

One can add a perturbation, $H_\text{in}$, to set the circuits inputs. We will write this as a projector onto the $n$ long binary bit vector $\textbf{x}$.  This 1-local projector has the form:
\begin{gather*}\label{eqn:binaryPro}
\Pi_{\textbf{x}}\bydef\ket{\textbf{x}^\perp}\bra{\textbf{x}^\perp}=\left(\frac{1}{2}\right)\sum_{i=1}^n\left(\openone +(-1)^{1-x_i} \sigma_i\right).
\end{gather*}
Now upper bound $\|H_\text{in}\|$ (for all two input and single output gates~\footnote{For the purpose of this section one is actually only concerned with the null space of the Hamiltonian and the spectral gap $\delta$ so $H_\text{prop}>\|  H_\text{in}\|$ is sufficient.}) as $\|H\|\leq 2$.  This implies that the spectral gap $\delta$ is greater than $2$.  By noticing that $\forall j, k$
\begin{gather*}
  H\ket{s_j}=\lambda\ket{s_j},\qquad   H\ket{s_k^\perp}=\lambda\ket{s_k^\perp}\\
\text{and}\qquad \bra{s_j}  H\ket{s_k^\perp} + \bra{s_k^\perp}  H\ket{s_j}=0,
\end{gather*}
 where $\ket{s_j}\in{\cal L}$ and $\ket{s_k}\in{\cal L^\perp}$, one recovers the strict equality, $\lambda(H_\text{in}|_{\cal L})=\lambda(H)$ (see Lemma~\ref{lemma:projection}).

Using combinations of these ground state logic gates, we will perform
computations.  For example, write the Hamiltonian with a low-energy
subspace in
\begin{equation}\label{eqn:span1}
\spn\{\ket{x_1 x_2}\ket{y_\star}\mid y_\star=x_1\wedge x_2, \forall x_1, x_2\in\{0,1\}\}
\end{equation}
as $H_{\wedge}(x_1,x_2,y_\star )$ and, with $y_\star$ defined in (\ref{eqn:span1}), write the Hamiltonian with a low-energy subspace in
\begin{equation*}
\spn\{\ket{x_3 y_\star}\ket{z_\star}\mid z_\star=y_\star\wedge x_2,\forall x_3\in\{0,1\}\}
\end{equation*}
as $H_{\wedge}(x_3,y_\star ,z_\star )$.  Then the proposition $x_1\wedge x_2\vee x_3=z_\star$ is constructed as a sum of terms:
\begin{equation}\label{eqn:e1}
H=\overbrace{H_{\vee}(x_1,x_2,y_\star)+H_{\wedge}(x_3,y_\star,z_\star)}^\text{$  H_\text{prop}$} +   H_\text{in}
\end{equation}
and the circuits input, $H_\text{in}$, is yet to be defined.  The qubit with label
$z_\star $ is now equal to $x_1\wedge
x_2\vee x_3$ and $y_\star $ is a temporary
variable that is equal to $x_1\wedge
x_2$, as seen in Table~\ref{table:gstt}.  

A small perturbation, $H_\text{in}$, can be added to set any of the qubits to specified values.  For example, to set the input as $x_1=1$, $x_2=0$ and $x_3=0$ one adds the perturbation $H_\text{in} = \ket{0}\bra{0}_1 + \ket{1}\bra{1}_2 + \ket{1}\bra{1}_3$.  If, alternatively, we were to let $H_\text{in}=\ket{0}\bra{0}_\star$, which acts on the circuits output $z_\star$, then the low-energy subspace would be spanned by all vectors where the output $z_\star $ is $\ket{1}$~\footnote{Assume that $H_\text{prop}$ represents a circuit and is given as an oracle Hamiltonian.  One wishes to search for an input bit string $x$ that will make the circuit output $z_\star=1$.  In this case, we will force an energy penalty any time the circuit outputs $0$ by acting on the output qubit, $z_\star$, with the Hamiltonian $H_\text{in}=\ket{0}\bra{0}_\star$.  After successful adiabatic evolution~\cite{FGGS00,FGG02}, qubits $x_1$, $x_2$ and $x_3$ can be measured to determine an input causing the circuit to output $1$.  If the circuit never outputs $1$, successful adiabatic evolution will return an input that \emph{minimizes} the Hamming distance from an input that would cause the circuit to output $1$.}.  As seen from Table~\ref{table:gstt}, this subspace is in
\begin{eqnarray*}
\spn\{\ket{001}\ket{1}\ket{0},\ket{011}\ket{1}\ket{0},\ket{101}\ket{1}\ket{0},\\
\ket{110}\ket{1}\ket{1},\ket{111}\ket{1}\ket{1}\},
\end{eqnarray*}
where we adhere to the ordering $\ket{x_1x_2x_3}\ket{z_\star}\ket{y_\star}$.  If instead we were to add the perturbation $H_\text{in}$ to the qubit labeled $\ket{y_\star}$, the ground space would be spanned by $\{\ket{110}\ket{1}\ket{1},\ket{111}\ket{1}\ket{1}\}$.

To complete our reduction, the 3-local Hamiltonians, just described, will be reduced in the next section to 2-local Hamiltonians. Before continuing to our 2-local reduction, let us state Lemma~\ref{lemma:projection} and Theorem~\ref{theorem:3local} --- the proof of which is implied by the results of this section.  Here we choose a finite set $\Omega$ of one-output Boolean functions as basis.  Then, an $\Omega$-circuit works for a fixed number of Boolean input variables and consists of a finite number of gates, where each gate is defined by it's type taken from $\Omega$.  (For additional background information on boolean functions and switching circuits see the freely available standard reference~\cite{Weg87}.)

\begin{theorem}\label{theorem:3local}
Let $f$ be a switching function given as the map $f:\{0,1\}^k \rightarrow\{0,1\}^m$ for finite $k$ and $m$.  Now let there be an asynchronous $\Omega$-circuit computing $f$.  Then there exists an $\Omega$-circuit embedding into the ground space of a 3-local Hamiltonian, $H_3$, such that: i.) The norm of the Hamiltonian $\|H_3\|$ is constant and, in particular independent of the size of $f$, the $\Omega$-circuit, as well as k and m. ii.) The $\Omega$-circuit embedding is upper bounded by a number of qubits $\mathcal{O}(1)$-reducible to the number of classical gates required on the same lattice.
\end{theorem}

An important technical tool used in our construction is a variant of the projection Lemma~\cite{KR03,AvDK+05,KKR06} --- proven in Appendix~\ref{appendix:projection}.  Let us denote $\mathcal{H}$ as a Hilbert space of interest and let $H_1$ be some Hamiltonian.  Consider a subspace ${\cal L}\in\mathcal{H}$ such that a Hamiltonian $H_2$ has the property that ${\cal L}$ is a $0$ eigenspace and ${\cal L}^\perp$ is an eigenspace of at least $\delta~(>2\|H_1\|)$.  Consider the Hamiltonian $H=H_1+H_2$, the projection lemma says that the lowest eigenvalue of $H$, $\lambda(H)$, is the lowest eigenvalue of $H_1$ restricted to the subspace ${\cal L}$ --- that is $\lambda(H_1|_{\cal L})$.  Thus, by adding $H_2$ one adds a penalty (proportional to $\delta$) to any vector in ${\cal L}^\perp$.  To state the Projection Lemma (Strict Equality) we:

\begin{lemma}\label{lemma:projection}
Let $H =H_1+H_2$ be the sum of two Hamiltonians operating on some Hilbert space $\mathcal{H} = {\cal L} + {\cal L}^\perp$.  Denote ${\cal L}=\spn\{\ket{s_j}|\forall j\}$ and ${\cal L}^\perp=\spn\{\ket{s_k^\perp}\forall k\}$ for finite $j,k$.  Consider the restriction $H_2|_{\cal L}=0$ and $H_2|_{{\cal L}^\perp}\geq\delta(>2\|H_1\|)$.  Then, if $\forall j,k$ $H\ket{s_j}=\lambda\ket{s_j}$ $(\forall k$ $H\ket{s_k^\perp}=\lambda\ket{s_k^\perp})$, $\bra{s_j}  H\ket{s_k^\perp} + \bra{s_k^\perp} H\ket{s_j}=0$ the following equality holds: $\lambda(H) = \lambda(H_1|_{\cal L})$.
\end{lemma}

\section{The 2-local reduction}\label{sec:2local}
The main result of this section can be found in Table~\ref{table:Htab}.  To develop this table we used the algebra of multi-linear forms~\cite{BH02} and the Karnaugh map method from discrete mathematics~\cite{Ros99} --- which we review in Appendix~\ref{appendix:kmap}.

We consider multi-linear forms that are maps $f$ from the Booleans numbers to the reals, where the inputs and outputs are of finite size.  For instance, the multi-linear form for \AND{} (\OR{}) is simply $f_\wedge=x_1\wedge x_2$~($f_\vee=x_1+x_2-2x_1\wedge x_2$).  Hence, one can express the Boolean equation $f=x_1\wedge x_2\vee x_3$ with the polynomial $f= x_1\wedge x_2+x_3-x_1\wedge x_2\wedge x_3$.  Let us first write the vector of integers:
\begin{equation}\label{eqn:c}
\textbf{c}^\intercal =(c_0,c_1,c_2,c_3,c_4,c_5,c_6,c_7,),
\end{equation}
representing the outputs of a multi-linear function $f$ over the three Boolean input arguments $x_1,x_2$ and $x_3$.  We wish to construct a canonical representation for any multi-linear function of three variables in terms of the vector $\textbf{c}$ from (\ref{eqn:c}).  We will represent the negation of the variable $x$ as $\bar x$ (or using the notational equivalent $\neg x$) and canonically expand (\ref{eqn:c}) as a sum of products:
\begin{eqnarray}\label{eqn:generic}
f(x_1,x_2,x_3)&=& c_0 \cdot{\bar x_1 \bar x_2 \bar x_3}+c_1 \cdot{\bar x_1\bar x_2 x_3}+c_2\cdot{\bar x_1 x_2 \bar x_3}\nonumber\\
&&+c_3 \cdot{\bar x_1 x_2 x_3}+c_4 \cdot{x_1 \bar x_2 \bar x_3}+c_5 \cdot{x_1 \bar x_2 x_3}\nonumber\\
&&+c_6 \cdot{x_1 x_2 \bar x_3}+c_7 \cdot{x_1x_2 x_3}.
\end{eqnarray}
This expansion (\ref{eqn:generic}) forms a basis for the space of 3-variable Hamiltonians, but to realize any of the eight terms requires 3-body couplings. This motivates us to write a second canonical expansion, found from a change of variables in (\ref{eqn:generic}) and by expanding each term into it's positive polarity form:
\begin{eqnarray}\label{eqn:generic1}
f(x_1,x_2,x_3)&=& a_0 + a_1\cdot{x_1}+a_2 \cdot{x_2}+a_3\cdot{x_3}\nonumber\\
&+&a_4 \cdot{x_1 x_2}+a_5 \cdot{x_1 x_3}+a_6 \cdot{x_2 x_3}\nonumber\\
&+&a_7 \cdot{x_1x_2 x_3}.
\end{eqnarray}
This equation (\ref{eqn:generic1}) also forms a basis for the space of realizable Hamiltonians of 3-spins.  In this suggestive form, however, we can truncate (\ref{eqn:generic1}) past $2^{nd}$ order and consider the subclass of Hamiltonians that can be realized by setting $a_7=0$.

Out of the 16 possible functions of 2-input and 1-output variable, it can be proven that only two are not realizable using 3-spins.  These are the 2-local penalty Hamiltonians for \XOR{} ($\oplus$) and \EQV{} ($\odot$)~\footnote{Where \emph{exclusive} \OR{} (\XOR{}) is given as $f_\oplus(x_1,x_2)\bydef x_1\oplus x_2=\bar x_1x_2\vee x_1\bar x_2= x_1+x_2-2x_1\wedge x_2$, and \emph{equivalence} (\EQV{}) as $f_\odot(x_1,x_2)\bydef x_1\odot x_2=\bar x_1\bar x_2\vee x_1 x_2= 1-x_1-x_2+2x_1\wedge x_2$.}, which are each possible to realize by adding a single mediator qubit (as seen in Table~\ref{table:Htab}).

We will explain in detail how the positive-semidefinite \AND{} penalty Hamiltonian, $H_\wedge$, is derived.  We anticipate that the details of our approach will aid others faced with Hamiltonian constructions.  Let $\cal L$ be the null space of $H_\wedge$ and let all higher eigenspaces be given as $\cal L^\perp$.  The penalty Hamiltonian has a null space, ${\cal L}$, spanned by the vectors $\{\ket{x_1 x_2}\ket{z_\star}|z_\star=x_1\wedge x_2, \forall x_1,x_2\in\{0,1\}\}$.  Denote $\delta$ as an energy penalty applied to any vector component in ${\cal L^\perp}$.  Our goal is to develop a Hamiltonian that adds a penalty of at least $\delta$ to any vector that does not satisfy the truth table of the $\AND$ gate --- that is, we want to add an energy penalty to any vector with a component that lies in ${\cal L^\perp}$.

\begin{figure}[t]
\begin{tabular}{lr}
  \begin{picture}(50,50)
{\Karnaughdiagram{3}{01234567}($z_\star $, $x_1$$x_2$)[a.)~~~]
\PrimImpl(30,5)(15,6) \PrimImpl(40,5)(15,8) \PrimImpl(35,10)(7,15) }
\end{picture} & ~~~~~~~~~~\begin{picture}(50,50)
{\Karnaughdiagram{3}{01234567}($z_\star $, $x_1$$x_2$)[b.)~~~]
\PrimImpl(30,9)(15,25) \PrimImpl(40,10)(16,22)
 \PrimImpl(30,5)(45,7)} %
\end{picture} \\
  \begin{picture}(50,50)
{\Karnaughdiagram{3}{000....0}($z_\star $, $x_1$$x_2$)[c.)~~~]
 \PrimImpl(30,5)(15,6) \PrimImpl(40,5)(15,8) \PrimImpl(35,10)(7,15)
\PrimImpl(30,5)(45,7) }
\end{picture} & ~
\end{tabular}\caption{Karnaugh maps: a.) 2-local (positive polarity) interactions circled (e.g. $q_1  x_1x_2+q_2  x_1\wedge z_\star +q_3  x_2\wedge f^\star $). b.) Linear (positive polarity) fields circled (e.g. $l_1  x_1+l_2  x_2+l_3  x_3$).  The interactions in cubes a.) and b.) form a basis for the space of realizable ($\leq$ 3 qubit) positive-definite logical gadget Hamiltonians expressible as: $H(x_1,x_2,z_\star )=k_0+l_1  x_1+l_2  x_2+l_3  x_3+q_1  x_1\wedge x_2+q_2  x_1\wedge z_\star +q_3  x_2\wedge z_\star $, where $\forall i$, $k_0, l_i, q_i \geq 0$.  c.) A Karnaugh map illustrating (with ovals) the linear and quadratic terms needed to set the null space of the Hamiltonian (\ref{eqn:Hwedge}) to be in $\spn\{\ket{x_1x_2}\ket{y_\star}|y_\star=x_1\wedge x_2, \forall x_1,x_2\in\{0,1\}\}$.}\label{fig:quad}
\end{figure}
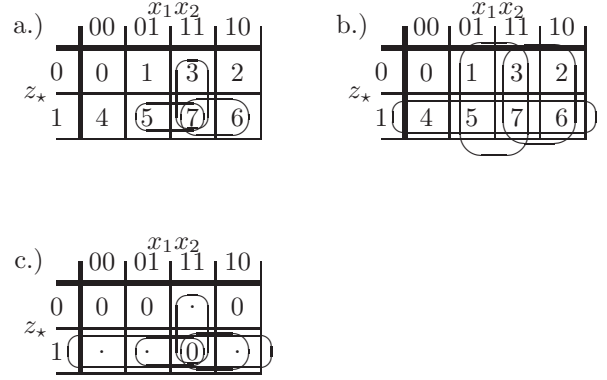

In order to make the penalty quadratic, one first constructs the Karnaugh map
illustrated in Fig.~\ref{fig:quad} c.) for the case $x_1\wedge x_2=z_\star$.  This is done by examining Table~\ref{table:NPH}.  In the right most column, all possible assignments for the variables $x_1$, $x_2$ and $z_\star$ are shown.  The Karnaugh map is constructed by examining the second column.  Whenever the variable $z_\star$ is not equal to the \AND{} of the variables $x_1$ and $x_2$, a penalty of at least $\delta$ must be applied, which ensures that vectors in the ground space satisfy $\ket{x_1}\ket{x_2}\ket{x_1\wedge x_2}$.  Any vector that must receive an energy penalty of $\delta$ is depicted in the Karnaugh map with a dot ($\cdot$).

Begin by noticing that any vector associated with cube number $4$ must receive an
energy penalty, so the 1-local field corresponding to the qubit with label $z_\star $ must be at least $\delta$ --- adding the term $p_1  z_\star$ to the Hamiltonian, with the constraint $p_1 \geq\delta$.  Cube $3$ must also receive an energy penalty of at
least $\delta$, adding the term $p_2   x_1\wedge  x_2$ to the
Hamiltonian $H_\wedge$.  With both penalties applied, vectors corresponding to cube $7$ must be brought back to the null space --- accomplished by subtracting the quadratic energy
rewards $r_1   z_\star\wedge x_1$ and $r_2   z_\star\wedge x_2$ from $H_\wedge$. A system
of equations for the Hamiltonian $H_\wedge(x_1,x_2,z_\star)=$
\begin{equation}\label{eqn:g1}
p_1   z_\star  + p_2   x_1\wedge x_2 - r_2   z_\star\wedge  x_1 - r_2  z_\star\wedge x_1
\end{equation}
can be solved to set the rewards ($r$'s) and the penalties ($p$'s). This system is derived from the
fact that the term $x_1x_2x_3$, corresponding to cube $7$, must have zero energy: $0=p_1+p_2-r_1-r_2$ and is
subject to the conditions that $p_1,p_2 \geq\delta$ and $|r_1+r_2|>p_1$.  For convenience, let $\delta = 1$ and then determine values for the coefficients in (\ref{eqn:g1}) and thus derive the 2-body Hamiltonian (for \AND{}):
\begin{equation}\label{eqn:Hwedge}
H_{\wedge}(x_1,x_2,z_\star)=3z_\star +x_1\wedge x_2-2z_\star\wedge x_1-2z_\star\wedge x_2.
\end{equation}

If one desires to invert an input variable, she simply
applies the transform: $\bar x_i \rightarrow (1-x_i)$.  For example, the
Hamiltonian applying the penalty $H_\wedge(\bar x_1,x_2,z_\star)$ is:
\begin{equation}\label{eqn:iso}
3z_\star +(1-x_1)\wedge x_2-2z_\star\wedge  (1-x_1)-2z_\star\wedge  x_2.
\end{equation}
To write this Hamiltonian in terms of spin variables, first change each variable, $x_i$, to its (matrix) operator form by the replacement $x_i\rightarrow\ket{0}\bra{0}_i$.  The change to spin variables is then accomplished by the replacement: $\ket{0}\bra{0}_i \rightarrow \frac{1}{2}(\openone-\sigma_i)$.  After these substitutions one arrives at the Hamiltonian (\ref{eqn:negH}) which is isomorphic ($\simeq$) to (\ref{eqn:iso}).
\begin{eqnarray}\label{eqn:negH}
&&H_\wedge(\bar x_1,x_2,z_\star)\simeq H_\wedge(-s_1,s_2,s_\star)=\\\nonumber
&=&\frac{1}{4}(3+\sigma_1-\sigma_2+2\sigma_\star-\sigma_1\sigma_2+2\sigma_1\sigma_\star-2\sigma_2\sigma_\star)
\end{eqnarray}

\begin{table}[t]
\small{\begin{tabular}{ccc||c||c}
  $x_1$ & $x_2$ & $z_\star $ & $z_\star \iseq x_1\wedge x_2$  & $H_\wedge(x_1,x_2,z_\star)$ \\\hline
  0 & 0 & 0 & $\bra{000}H_\wedge\ket{000}=0$  & 0 \\
  0 & 0 & 1 & $\bra{001}H_\wedge\ket{001}\geq\delta$ & $3 \delta$\\
  0 & 1 & 0 & $\bra{010}H_\wedge\ket{010}=0$ & 0 \\
  0 & 1 & 1 & $\bra{011}H_\wedge\ket{011}\geq\delta$ & $\delta$\\
  1 & 0 & 0 & $\bra{100}H_\wedge\ket{100}=0$ & 0 \\
  1 & 0 & 1 & $\bra{101}H_\wedge\ket{101}\geq\delta$ & $\delta$ \\
  1 & 1 & 0 & $\bra{110}H_\wedge\ket{110}\geq\delta$ & $\delta$ \\
  1 & 1 & 1 & $\bra{111}H_\wedge\ket{111}=0$ & 0 \\
\end{tabular}}\caption{Left column: possible assignments of the variables $x_1$, $x_2$ and $z_\star$.  Center column: illustrates the variable assignments that must receive an energy penalty $\geq\delta$.  Right column: truth table for $H_\wedge(x_1,x_2,z_\star )=3z_\star +x_1\wedge x_2-2z_\star \wedge x_1-2z_\star \wedge x_2$, which has a null space ${\cal L}\in\spn\{\ket{x_1 x_2}\ket{z_\star}|z_\star=x_1\wedge x_2,\forall x_1,x_2\in\{0,1\}\}$.}\label{table:NPH}%
\end{table}

We now have the necessary machinery in place to state two theorems (\ref{theorem:klocalboolean} and \ref{theorem:2localcir}).  In the first, we are concerned with a situation that arises in several applications.  That is, one often needs to couple three Boolean variables (\AND{} product), as $x_1\wedge x_2\wedge x_3$, using
only 2-local Hamiltonians.  From our reduction, it is possible to efficiently construct any k-local product term, $x_1\wedge x_2\wedge \cdots \wedge x_k$, of this type.  We prove this in Theorem~\ref{theorem:klocalboolean}.  We then present Theorem~\ref{theorem:2localcir}, which is a 2-local variant of Theorem~\ref{theorem:3local} --- the proof of which follows directly from the results of this section.


\begin{theorem}\label{theorem:klocalboolean}
Let $f_k$ be a k-local multi-linear from and let there be a Hamiltonian $H_k$ acting on the Hilbert space $\mathcal{H}_k$ such that $f_k \simeq H_k$.  Then there exists a 2-local multi-linear form, $f_2$, and corresponding Hamiltonian, $H_2$, acting on the Hilbert space $\mathcal{H}_2$ (where $\mathcal{H}_k\subseteq\mathcal{H}_2$), with the same low-energy subspace of $H_2$ in $\spn\{\ket{x}\ket{y}|y=f_k(x), \forall x\in\{0,1\}^n, \forall y\in\{0,1\}^m\}\subseteq\mathcal{H}$.  The number of mediator qubits required to realize $H_2$ is upper bounded by $\mathcal{O}(\text{size}(f_k))$.  In addition, the spectral gap of $H_2$ is bounded by the spectral gap of $H_k$.
\end{theorem}

\begin{proof}
To construct such a Hamiltonian, we will employ an inductive argument and consider a single (out of $w$) k-local term, $h_k=x_1\wedge x_2\wedge \cdots \wedge x_k$, that couples $k\geq 3$ Boolean variables.  We will now show the existence of a 2-local reduction requiring $(k-2)$ mediator qubits to embed $h_k$ into the ground state of a 2-local Hamiltonian.  Consider the 2-local coupling $z_\star\wedge  x_3$ and
add the Hamiltonian that forces an energy penalty whenever $z_\star$ is not the Boolean
\AND{} of the variables $x_1$ and $x_2$.  The 2-local Hamiltonian is written as
\begin{eqnarray*}
H_\wedge(x_1,x_2,z_\star)+z_\star\wedge x_3&\simeq&\frac{1}{4}(4-\sigma_1-\sigma_2+\\
3\sigma_\star+\sigma_3+\sigma_1\sigma_2&-&2\sigma_1\sigma_\star-2\sigma_2\sigma_\star+\sigma_\star\sigma_3),
\end{eqnarray*}
where $H_\wedge$ is found in Table~\ref{table:Htab}, and $z_\star$ is a temporary variable.  In words, the variable $z_\star$ is coupled to $x_3$ and the penalty, $H_\wedge$, forces $z_\star$ to be equal to the Boolean product of $x_1$ and $x_2$ --- thereby creating the desired coupling with respect to the subspace spanned by $\ket{x_1x_2x_3}$, $\forall i\in\{1,2,3\}, x_i\in\{0,1\}$.  For a k-local term $x_1\wedge x_2\wedge \cdots \wedge x_k$, this procedure is recursively repeated $k-2$ times.  The reduction requires $w(k-2)$ qubits to capture the low-lying eigenspace of $H_k$ with $H_2$.
\end{proof}

\begin{theorem}\label{theorem:2localcir}
Let $f$ be a switching function with a fixed number of inputs k and outputs m.  Let there be an asynchronous $\Omega$-circuit computing $f$ over the basis $\{\neg, \oplus, \wedge\}$.  There exists an $\Omega$-circuit embedding into the ground state of a 2-local Hamiltonian, $H_2$, such that: i.) The norm of the Hamiltonian $\|H_2\|$ is constant and, in particular independent of the size of $f$, the $\Omega$-circuit, k as well as m. ii.) The $\Omega$-circuit embedding is upper bounded by a number of qubits $\mathcal{O}(k)$-reducible to the number of classical gates required on the same lattice.
\end{theorem}

\section{A novel 3-local gadget}\label{sec:novelgadget}
We are concerned with constructing the ground state of the operator $J\sigma_1\otimes \sigma_2 \otimes\sigma_3$ --- which is a different task than coupling (that is, the \AND{} product) three Boolean variables $x_1\wedge x_2\wedge x_3$.  Without loss of generality, let us consider construction of the target Hamiltonian
\begin{equation}\label{eqn:Ht}
H_\text{target} = Y + J\sigma_1\otimes\sigma_2\otimes \sigma_3,
\end{equation}
where $Y$ is diagonal in the $\sigma$ basis.  We will write the spectrum of $\sigma_1\otimes \sigma_2\otimes \sigma_3$, in canonical (Boolean counting) order, as $\{1,-1,-1,1,-1,1,1,-1\}$~\footnote{This spectrum corresponds to the Walsh function represented by the $8^{th}$ column of the matrix ${\sc H}^{\otimes 3}$, where ${\sc H}$ is the $2\times2$ Hardamard matrix.  We remark that $\{\{0,1\}, \oplus, \wedge\}$ is the Galois field $\mathbb{Z}_2$.}.  Now the low-energy, $\lambda(\sigma_1\otimes \sigma_2\otimes \sigma_3)=-1$, eigenspace is given as
\begin{equation*}
{\cal L} =\spn\{\ket{001}, \ket{010}, \ket{100}, \ket{111}\}
\end{equation*}
and the high-energy, $+1$, eigenspace as
\begin{equation*}
{\cal L^\perp} =\spn\{\ket{000}, \ket{011}, \ket{101}, \ket{110}\}.
\end{equation*}
Over the complex field, the tensor product ($\otimes$) of two elements is simply their complex multiplicative ($\cdot$) product.  With respect to the canonical order, the spin variables for this operator (\ref{eqn:Ht}) form the product $z_\star=s_1\cdot s_2\cdot s_3,~~\text{where}~~\forall i, s_i\in\{+1,-1\}$, and so we consider the group Homomorphism $\{-1,+1,\cdot\}\mapsto \{1,0,\oplus\}$, where $\oplus$ denotes modulo 2 sum (\XOR{}); whence
\begin{equation*}
z_\star = x_1\oplus x_2\oplus x_3,~\forall x_1,x_2,x_3\in\{0,1\}^3.
\end{equation*}
In what follows, we will present a general framework to construct the ground state of any operator in the $\sigma$ basis and apply this approach to produce a 3-local gadget requiring three mediator qubits.  We will then focus our attention on optimization of this new 3-local gadget, which is shown to be possible to realize using only two mediator qubits.

Let us state an overview of our approach.  To capture both the low- and high-energy spectrum, while preserving the spectral gap, one will first write down a penalty Hamiltonian for the 3-variable function $z_\star$, which acts on the Hilbert space $\mathcal{H}$.  This function, $z_\star$, outputs logical 0 for any input vector in ${\cal L}$, and for all vectors in ${\cal L^\perp}$ the function outputs logical 1. We will next add a small perturbation to the output $z_\star$ --- thereby breaking the low-energy degeneracy and allowing us to capture the spectrum of (\ref{eqn:Ht}) exactly, with respect to the subspace
\begin{equation*}
\mathcal{L+L^\perp}=\spn\{\ket{x_1x_2x_3}|\forall x_1,x_2,x_3\in\{0,1\}\}\subset\mathcal{H}.
\end{equation*}

\textbf{3-local gadgets with 3 mediator qubits:} From Table~\ref{table:Htab} we know that each \XOR{} function requires an extra qubit, and so three mediator qubits are required to create the desired coupling.  Let us write the Hamiltonian that applies the \XOR{} penalty to the variables $x_1$ and $x_2$ as $H_\oplus(x_1,x_2,y_\star,m_1)$ and the Hamiltonian that applies the \XOR{} penalty to the variables $x_3$ and $y_\star$ as $H_\oplus(x_3,y_\star,z_\star, m_2)$.  Now order the variables as $\ket{x_1x_2x_3}\ket{z_\star}\ket{y_\star m_1m_2}$, where $m_1$ and $m_2$ are mediator qubits and $y_\star$ is a temporary variable that is not read.  To split the spectrum into it's respective low-energy (${\cal L}$) and high-energy (${\cal L^\perp}$) subspaces we add the perturbation $V= J(\ket{0}\bra{0} - \ket{1}\bra{1})$, which acts on the qubit ${z_\star}$.  This allows one to construct the Hamiltonian (\ref{eqn:Ht}), with the desired spectrum since the commutator $[Y,J\sigma_1\otimes\sigma_2 \otimes\sigma_3]=0$ shows that $Y$ only adds energy shifts and not level mixing --- see Lemma~\ref{lemma:projection}.

\textbf{3-local gadgets with 2 mediator qubits:} Let us present an alternative approach to realizing a 3-local gadget which requires only two mediator qubits.  To construct the gadget Hamiltonian, consider the 2-local coupling $z_4^\star s_3$ and add the Hamiltonian that forces a penalty whenever $z_4^\star$ is not equal to the \EQV{} of variables $x_2$ and $x_3$.  The 2-local Hamiltonian is written as $H_\oplus(x_1,x_2,z_4^\star)+z_4^\star s_3$, where $H_\oplus$ is found in Table~\ref{table:Htab}, and $z_4^\star$ is a temporary variable.  The Hamiltonian (\ref{eqn:gadget}) captures the desired spectrum for $\delta>2|J|$.
\begin{eqnarray}\label{eqn:gadget}
H &=& \frac{\delta}{2} (4 + \sigma_2 \sigma_3 +(\sigma_2 + \sigma_3)\sigma_4 +\\\nonumber
 &&2 (\openone - \sigma_2 -\sigma_3 - \sigma_4)\sigma_5 - \sigma_2 -\sigma_3 -\sigma_4  ) + \underbrace{J\sigma_1 \sigma_4}_\text{$s_1z_4^\star$}.
\end{eqnarray}
The ground space of the Hamiltonian (\ref{eqn:gadget}) is given as
\begin{equation*}
{\cal L}= \spn\{\ket{001}\ket{00},\ket{010}\ket{11},\ket{100}\ket{11},\ket{111}\ket{01}\}
\end{equation*}
and the first excited space as
\begin{equation*}
{\cal L^\perp} =\spn\{\ket{000}\ket{01},\ket{100}\ket{00},\ket{110}\ket{00},\ket{110}\ket{10}\},
\end{equation*}
where the qubits are in ascending order: qubit 4 represents the Boolean \EQV{} of qubits 2 and 3, while qubit 5 is the mediator qubit needed to construct the \EQV{} function.

We will now state then prove Theorem~\ref{theorem:manylevels} which allows one to construct, not only the ground state, but several levels of the low-lying energy subspace of k-body interactions using only 2-body Hamiltonians, formally we

\begin{theorem}\label{theorem:manylevels}
Let $H_k$ be a k-local Hamiltonian diagonal in any basis $\sigma$ and let this Hamiltonian act on the Hilbert space $\mathcal{H}_k$.  Assert that $H_k$ has bounded norm, and let the strictly increasing list $\{E_1,E_2,\cdots,E_k\}$ denote the eigenenergies of $H_k$ formed by combing degeneracies, and label the corresponding eigenspaces as $\{{\cal L}_1,{\cal L}_2,\cdots,{\cal L}_k\}$, respectively.  Then there exists a 2-local Hamiltonian, $H_2$, with a low-lying spectrum isomorphic to that of $H_k$.  Moreover, $H_2$ is equivalent to $H_k$ with respect to a subspace spanned by $\{{\cal L}_1,{\cal L}_2,\cdots,{\cal L}_k\}$.  In particular, there exists a 2-local reduction capturing the $k$ energy subspaces $\{{\cal L}_1,{\cal L}_2,\cdots,{\cal L}_k\}$ in the low-energy subspace $H_k$.
\end{theorem}

\begin{proof}
Let us review the general method to construct ground states.  First, determine ${\cal L}$, the low-energy subspace, and let $E_g$ denote the ground state energy.  One will next write a function, $z_\star=f(x_1,x_2,\dots,x_n)$, that outputs 0 for all input vectors in ${\cal L}$, and for all other vectors the function will output 1.  The ground state will be realized with respect to a subspace spanned by the qubits labeled $\ket{x_1}\ket{x_2},\dots,\ket{x_n}$.  To capture the desired ground space, a perturbation ($V=E_g\ket{0}\bra{0}$) is added, which only acts on the qubit ${z_\star}$.  Assume that we are instead interested in capturing several energy subspaces, with energies $\{E_1,E_2,\cdots,E_k\}$, and let us label these spaces as $\{{\cal L}_1,{\cal L}_2,\cdots,{\cal L}_k\}$, respectively.  We will construct a function with $k$ outputs, and repeat the process outlined above --- this time acting on each respective $j^{th}$ function with the perturbation $V=\sum_{j=1}^kE_j\ket{0}\bra{0}_j$.
\end{proof}


\section{Conclusion}\label{sec:con}

In this study, we have adapted a range of classical algebraic reduction methods to the construction of the low-lying energy subspace of k-local Hamiltonians using 2-local Hamiltonians.  Our methods do not rely on perturbation theory or the associated large spectral gap.  We have examined explicit constructions of various useful k-local to 2-local conversion Hamiltonians --- including both those needed to embed logical functions as well as couple spin variables.  We have found constructions of these Hamiltonians which are optimal in the number of introduced ancillary qubits.  For ease of reference, our results are summarized in Table~\ref{table:Htab} and Table~\ref{table:gadgetTab}.  In Theorem~\ref{theorem:manylevels} we presented a novel method to construct several levels, including the ground state, of the low-lying energy subspace of k-body interactions using 2-body Hamiltonians.   Our methods have several applications in adiabatic quantum algorithm design and quantum complexity theory.


\section{Acknowledgments}
I thank Peter J. Love, Stephen Jordan and Barbara Terhal.  This work received funding from EC FP6 STREP QICS and the Faculty of Arts and Sciences at Harvard University.

\begin{table*}[f]
\small{\begin{tabular}{c||c||c}
  $\ket{x_1x_2x_3}$ & $\ket{z_\star}$ & $\ket{y_\star}$\\\hline
  $\ket{000}$ & $\ket{0}$ & $\ket{0}$ \\
  $\ket{001}$ & $\ket{1}$ & $\ket{0}$ \\
  $\ket{010}$ & $\ket{0}$ & $\ket{0}$ \\
  $\ket{011}$ & $\ket{1}$ & $\ket{0}$ \\
  $\ket{100}$ & $\ket{0}$ & $\ket{0}$ \\
  $\ket{101}$ & $\ket{1}$ & $\ket{0}$ \\
  $\ket{110}$ & $\ket{1}$ & $\ket{1}$ \\
  $\ket{111}$ & $\ket{1}$ & $\ket{1}$ \\
\end{tabular}}\caption{Ground state truth table generated for the Hamiltonian (\ref{eqn:e1}).  The function output, $z_\star$, is equal to $x_1\wedge x_2\vee x_3$.  It is instructive to think of the variable $y_\star$ as a \emph{coupler} that follows the variables $x_1$ and $x_2$ as $y_\star=x_1\wedge x_2$.}\label{table:gstt}%
\end{table*}

\begin{table*}[f]
{\small\center \begin{tabular}{c||c||c}
  function & 2-local Hamiltonian $\text{H}(x_1,x_2,z_\star )=$ & ground state~(ordered: $\ket{x_1}\ket{x_2}\ket{z_\star}$) \\\hline
  $0 = z_\star$ & $\frac{1}{2}(\openone -\sigma_{3})$ & $\spn\{\ket{x_1x_2}\ket{0}|\forall x_1,x_2\in\{0,1\}\}$ \\
  $1= z_\star$ & $\frac{1}{2}(\openone +\sigma_{3})$ & $\spn\{\ket{x_1x_2}\ket{1}|\forall x_1,x_2\in\{0,1\}\}$ \\\hline\hline
  $\bar x_1 \wedge \bar x_2= z_\star$ & $\frac{1}{4}(3+\sigma_1+\sigma_2-2\sigma_\star+\sigma_1\sigma_2-2\sigma_1\sigma_\star-2\sigma_2\sigma_\star)$ & $\spn\{\ket{001},\ket{010},\ket{100},\ket{110}\}$ \\
  $\bar x_1 \wedge  x_2= z_\star$ & $\frac{1}{4}(3+\sigma_1-\sigma_2+2\sigma_\star-\sigma_1\sigma_2+2\sigma_1\sigma_\star-2\sigma_2\sigma_\star)$ & $\spn\{\ket{000},\ket{011},\ket{100},\ket{110}\}$ \\
  $x_1 \wedge x_2= z_\star$ & $\frac{1}{4}(3-\sigma_1-\sigma_2+2\sigma_\star+\sigma_1\sigma_2-2\sigma_1\sigma_\star-2\sigma_2\sigma_\star)$ & $\spn\{\ket{000},\ket{010},\ket{100},\ket{111}\}$ \\
  $x_1 \wedge \bar x_2= z_\star$ & $\frac{1}{4}(3-\sigma_1+\sigma_2+2\sigma_\star-\sigma_1\sigma_2-2\sigma_1\sigma_\star+2\sigma_2\sigma_\star)$ & $\spn\{\ket{000},\ket{010},\ket{101},\ket{110}\}$ \\\hline\hline
   $x_1 \vee x_2= z_\star$ & $\frac{1}{4}(4+\sigma_1+\sigma_2-2\sigma_\star+2\sigma_1\sigma_2-3\sigma_1\sigma_\star-3\sigma_2\sigma_\star)$ &
   $\spn\{\ket{000},\ket{011},\ket{101},\ket{111}\}$ \\
   $x_1 \vee \bar x_2= z_\star$ & $\frac{1}{4}(4+\sigma_1-\sigma_2-2\sigma_\star-2\sigma_1\sigma_2-3\sigma_1\sigma_\star+3\sigma_2\sigma_\star)$  & $\spn\{\ket{001},\ket{010},\ket{101},\ket{111}\}$ \\
   $\bar x_1 \vee \bar x_2= z_\star$ & $\frac{1}{4}(4-\sigma_1-\sigma_2+2\sigma_\star+2\sigma_1\sigma_2-3\sigma_1\sigma_\star-3\sigma_2\sigma_\star)$  &
    $\spn\{\ket{001},\ket{011},\ket{101},\ket{110}\}$  \\
   $\bar x_1 \vee x_2= z_\star$ & $\frac{1}{4}(4-\sigma_1+\sigma_2-2\sigma_\star-2\sigma_1\sigma_2+3\sigma_1\sigma_\star-3\sigma_2\sigma_\star)$  &
    $\spn\{\ket{001},\ket{011},\ket{100},\ket{111}\}$  \\\hline\hline
   $x_1 \nLeftrightarrow z_\star$ & $\frac{1}{2}(\openone +\sigma_{1}\sigma_{3})$ & $\spn\{\ket{0x_21},\ket{1x_20}|\forall x_2\in\{0,1\}\}$ \\
   $x_2\Leftrightarrow z_\star$ & $\frac{1}{2}(\openone -\sigma_{2}\sigma_{3})$ &  $\spn\{\ket{x_100},\ket{x_111}|\forall x_1\in\{0,1\}\}$\\
   $x_1\Leftrightarrow z_\star$ & $\frac{1}{2}(\openone-\sigma_{1}\sigma_{3})$ & $\spn\{\ket{0x_20},\ket{1x_21}|\forall x_2\in\{0,1\}\}$ \\
   $x_2 \nLeftrightarrow z_\star$ & $\frac{1}{2}(\openone+\sigma_{2}\sigma_{3})$ & $\spn\{\ket{x_101},\ket{x_110}|\forall x_1\in\{0,1\}\}$ \\\hline\hline
   $x_1\oplus x_2=z_\star$ &  $4+\sigma_1\sigma_2 + (\sigma_1+\sigma_2)\sigma_\star +
2( \openone-\sigma_1-\sigma_2-\sigma_\star)\sigma_4 - \sigma_2 - \sigma_\star- \sigma_4$ & $\spn\{\ket{0000},\ket{0111},\ket{1011},\ket{1101}\}$ \\
   $x_1\odot x_2=z_\star$  & $4-\sigma_1\sigma_2 + (\sigma_1-\sigma_2)\sigma_\star +
2(\openone-\sigma_1+\sigma_2-\sigma_\star)\sigma_4 + \sigma_2 - \sigma_\star- \sigma_4$  & $\spn\{\ket{0100},\ket{0011},\ket{1111},\ket{1001}\}$ \\
\end{tabular}}
\caption{Logical gadgets ($\S$~\ref{sec:2local}): The span of the zero energy ground space (${\cal{L}}$) of these gadget Hamiltonians represent the truth table of a given switching function in the spin variables (as, for instance, the \AND{} function: ${\cal{L}}=\spn\{\ket{x_1x_2}\ket{z_\star }|z_\star = x_1\wedge x_2,\forall x_1,x_2\in\{0,1\}\}$).  This table includes all $16=2^{2^n}$ possible boolean functions with $n=2$ inputs.}\label{table:Htab}
\end{table*}

\begin{table*}[f]
{\small\center
\begin{tabular}{c||c}
  3-local coupling &  2-local Hamiltonian \\\hline
  $Jx_1\wedge x_2\wedge x_3$ &$\frac{1}{4}(4-\sigma_1-\sigma_2+3\sigma_\star+\sigma_3+\sigma_1\sigma_2-2\sigma_1\sigma_\star-2\sigma_2\sigma_\star+J\sigma_\star\sigma_3)$\\
   \hline\hline
  $J\sigma_1\otimes\sigma_2\otimes\sigma_3$ & $\frac{\delta}{2} (4 + \sigma_2 \sigma_3 +(\sigma_2 + \sigma_3)\sigma_4 + 2 (\openone - \sigma_2 -\sigma_3 - \sigma_4)\sigma_5 - \sigma_2 -\sigma_3 -\sigma_4  ) + J\sigma_1 \sigma_4$ \\

\end{tabular}
}
\caption{3-local gadgets.  Top ($\S$~\ref{sec:2local}): Hamiltonian with a low-energy subspace that couples three Boolean variables.  The state of the mediator qubit $\sigma_\star$ is a function (the \AND{}) of qubits 1 and 2.  Bottom ($\S$~\ref{sec:novelgadget}): Hamiltonian with low-energy subspace that couples three spin variables for $\delta>2|J|$.  The ground space, ${\cal L}=\spn\{\ket{001}\ket{00},\ket{010}\ket{11},\ket{100}\ket{11},\ket{111}\ket{01}\}$ and the first excited space, ${\cal L^\perp}=\spn\{\ket{000}\ket{01},\ket{100}\ket{00},\ket{110}\ket{00},\ket{110}\ket{10}\}$ --- the qubits are in ascending order: qubit 4 represents the Boolean \EQV{} of qubits 2 and 3, while qubit 5 is the mediator qubit needed to construct the \EQV{} function.}\label{table:gadgetTab}
\end{table*}



\appendix

\section{Projection Lemma}\label{appendix:projection}

We will now prove Lemma~\ref{lemma:projection} which is discussed on page 3 in $\S$~\ref{sec:3local}.  Consider first the case that $\lambda(H)\leq\lambda(  H_1|_{\cal L})$.  Denote by $\ket{\eta}\in{\cal L}$ the minimizing eigenvector of $  H_1|_{\cal L}$ with eigenvalue $\lambda(H_1|_{\cal L})$.  Since $  H_2\ket{\eta} = 0$,
\begin{equation*}
\bra{\eta}  H\ket{\eta} = \bra{\eta}  H_1\ket{\eta}+\bra{\eta}  H_1\ket{\eta} = \lambda(H_1|_{\cal L}).
\end{equation*}
Now consider actually minimizing over all vectors $\ket{\zeta}$ of unit length:
\begin{equation*}
\min_{\ket{\zeta}\in~{\cal L} + {\cal L}^\perp }\{\bra{\zeta}  H\ket{\zeta}\}\leq\bra{\eta}  H\ket{\eta}=\lambda(H_1|_{\cal L}),
\end{equation*}
proving a R.H.S. To show the lower bound on $\lambda(H)$ write any unit vector $\ket{v}\in \mathcal{H}={\cal L}+{\cal L}^\perp$ as $\ket{v} = \alpha\ket{s} + \beta\ket{s^\perp}$ where $\ket{s}$ ($\ket{s^\perp}$) is in ${\cal L}$ (${\cal L^\perp}$), $\alpha,\beta\in\mathbb{R}$, $\alpha,\beta\geq0$ and $\alpha^2+\beta^2=1$.  So
\begin{eqnarray*}
\lambda(H)&=&\lambda(H_1+  H_2)\geq \alpha^2\bra{s}  H_1\ket{s} + \alpha\beta(\bra{s}  H_1\ket{s^\perp} + \\\nonumber
&&\bra{s^\perp}  H_1\ket{s}) + \beta^2\bra{s^\perp}  H_1\ket{s^\perp} + \delta\beta^2.
\end{eqnarray*}
For real $  H_1$, $\ket{\psi}$ and $\ket{\phi}$:
\begin{eqnarray*}
\bra{\psi}  H_1\ket{\phi}&=&\bra{\psi}  H_1\ket{\phi} \Rightarrow \\
&& \alpha\beta(\bra{s}  H_1\ket{s^\perp} + \bra{s^\perp}  H_1\ket{s})=2\alpha\beta\bra{s}  H_1\ket{s^\perp}.
\end{eqnarray*}
However, $\ket{s}$ and $\ket{s^\perp}$ are eigenstates of $H_1$ and $\bra{s}s^\perp\rangle=0$, hence:
\begin{equation*}
\lambda(H_1+H_2)\geq \lambda(H_1|_{\cal L}) +\beta^2(\delta-2\|  H_1\|)
\end{equation*}
is minimized with $\beta=0$ so the projection lemma becomes
\begin{equation*}
\lambda(  H_1|_{\cal L})\leq\lambda(H)\leq\lambda(H_1|_{\cal L})~\Rightarrow~\lambda(H)=\lambda(H_1|_{\cal L}).~\Box
\end{equation*}

\section{Karnaugh maps}\label{appendix:kmap}

The \emph{Karnaugh map} is a tool to facilitate the algebraic reduction of Boolean functions.  We made use of this tool in $\S$~\ref{sec:2local} during explanation of the specific details required to construct Tables~\ref{table:Htab} and~\ref{table:gadgetTab}.  Many excellent texts and online tutorials cover the use of Karnaugh maps such as the wikipedia entry (\href{http://en.wikipedia.org/wiki/Karnaugh_map}{\tt http://en.wikipedia.org}), the articles linked to therein as well as the straight forward reference~\cite{Ros99}.  This Appendix briefly introduces these maps to make the present paper self contained.

Karnaugh maps (see Fig~\ref{fig:quad} for three examples), or more compactly K-maps, are organized so that the truth table of a given equation, such as a Boolean equation ($f:\mathbb{B}^n\rightarrow \mathbb{B}$) or multi-linear form ($f:\mathbb{B}^n\rightarrow \mathbb{R}$), is arranged in a grid form and between any two adjacent boxes only one domain variable can change value.

This ordering results as the rows and columns are ordered according to Gray code --- a binary numeral system where two successive values differ in only one digit.  For example, the 4-bit Gray code is given as:
\begin{eqnarray*}
&&\{0000, 0001, 0011, 0010, 0110, 0111, 0101, 0100, 1100,\\
&&1101, 1111, 1110, 1010, 1011, 1001, 1000\}.
\end{eqnarray*}
By arranging the truth table of a given function in this way, a K-map can be used to derive a minimized function.

To use a K-map to minimize a Boolean function one \emph{covers} the 1's on the map by rectangular \emph{coverings} containing a number of boxes equal to a power of 2.  For example, one could circle a map of size $2^n$ for any constant function $f=1$.  Fig~\ref{fig:quad} a.) and b.) contain three circles each --- all of 2 and 4 boxes respectively. After the 1's are covered, a term in a \emph{sum of products expression}~\cite{Weg87} is produced by finding the variables that do not change throughout the entire covering, and taking a 1 to mean that variable ($x_i$) and a 0 as its negation ($\overline{x_i}$). Doing this for every covering yields a function which \emph{matches} the truth table.

For instance consider Fig~\ref{fig:quad} a.) and b.).  Here the boxes contain simply labels representing the decimal value of the corresponding Gray code ordering.  The circling in Fig~\ref{fig:quad} a.) would correspond to the truth vector (ordered $z_\star, x_1$ then $x_2$)
\begin{equation}\label{eqn:tt1}
\left(0,0,0,1,0,1,1,1\right)^T.
\end{equation}
The cubes 3 and 7 circled in Fig~\ref{fig:quad} correspond to the sum of products term $x_1x_2$.  Likewise (5,7) corresponds to $z_\star x_2$ and finally (7,6) corresponds to $z_\star x_1$. The sum of products representation of (\ref{eqn:tt1}) is simply
\begin{equation*}
f(z_\star,x_1,x_2)=x_1x_2\vee z_\star x_2\vee z_\star x_1.
\end{equation*}
Let us repeat the same procedure for Fig~\ref{fig:quad} b.) by again assuming the circled cubes correspond to 1's in the functions truth table.  In this case one finds $z_\star$ for the circling of cubes ladled (4,5,7,6), $x_2$ for (1,3,5,7) and $x_1$ for (3,2,7,6) resulting in the function
\begin{equation*}
f(z_\star,x_1,x_2)=x_1\vee z_\star \vee x_2.
\end{equation*}

Our use of K-maps in $\S$~\ref{sec:2local} allows one to visualize cube groups (variable products) that are at most 2-local in size --- the highest order terms realizable with 2-local Hamiltonians.   In addition, K-maps help reduce the number of simultaneous equations that, as seen in $\S$~\ref{sec:2local}, must be solved --- see (\ref{eqn:g1}) and (\ref{eqn:Hwedge}).  The Karnaugh maps shown in Fig.~\ref{fig:quad} a.) and b.) illustrate groupings for quadratic and linear
interactions, respectively corresponding to 2-body terms and 1-local fields.  In $\S$~\ref{sec:2local}, this observation allowed us to derive 2-local Hamiltonians and prove which Hamiltonians are not possible to construct given specific numbers of mediator qubits.

\end{document}